\documentclass[12pt,epsf,psfig]{article}
\usepackage{graphicx}

\begin{document}


\def\a{\alpha}
\def\b{\beta}
\def\c{\varepsilon}
\def\d{\delta}
\def\e{\epsilon}
\def\f{\phi}
\def\g{\gamma}
\def\h{\theta}
\def\k{\kappa}
\def\l{\lambda}
\def\m{\mu}
\def\n{\nu}
\def\p{\psi}
\def\q{\partial}
\def\r{\rho}
\def\s{\sigma}
\def\t{\tau}
\def\u{\upsilon}
\def\v{\varphi}
\def\w{\omega}
\def\x{\xi}
\def\y{\eta}
\def\z{\zeta}
\def\D{\Delta}
\def\G{\Gamma}
\def\H{\Theta}
\def\L{\Lambda}
\def\F{\Phi}
\def\P{\Psi}
\def\S{\Sigma}
\def\V{\varPsi}

\def\o{\over}
\newcommand{\sla}[1]{#1 \llap{\, /}}

\newcommand{\beq}{\begin{eqnarray}}
\newcommand{\eeq}{\end{eqnarray}}
\newcommand{\gsim}{ \mathop{}_{\textstyle \sim}^{\textstyle >} }
\newcommand{\lsim}{ \mathop{}_{\textstyle \sim}^{\textstyle <} }
\newcommand{\vev}[1]{ \left\langle {#1} \right\rangle }
\newcommand{\bra}[1]{ \langle {#1} | }
\newcommand{\ket}[1]{ | {#1} \rangle }
\newcommand{\EV}{ {\rm eV} }
\newcommand{\KEV}{ {\rm keV} }
\newcommand{\MEV}{ {\rm MeV} }
\newcommand{\GEV}{ {\rm GeV} }
\newcommand{\TEV}{ {\rm TeV} }
\def\diag{\mathop{\rm diag}\nolimits}
\def\Spin{\mathop{\rm Spin}}
\def\SO{\mathop{\rm SO}}
\def\O{\mathop{\rm O}}
\def\SU{\mathop{\rm SU}}
\def\U{\mathop{\rm U}}
\def\Sp{\mathop{\rm Sp}}
\def\SL{\mathop{\rm SL}}
\def\tr{\mathop{\rm tr}}


\baselineskip 0.7cm

\begin{titlepage}

\begin{flushright}
UT-04-25 \\
\end{flushright}

\vskip 1.35cm
\begin{center}
{\Large \bf
Realization of Minimal Supergravity
}
\vskip 1.0cm
M.~Ibe, Izawa K.-I., and T.~Yanagida
\vskip 0.4cm

{\it Department of Physics, University of Tokyo,\\
     Tokyo 113-0033, Japan}

\vskip 1.5cm

\abstract{
Minimal supergravity mediation of supersymmetry breaking has attracted
 much attention due to its simplicity, which leads to its predictive
 power.  
We consider how Nature possibly realizes minimal supergravity through
 inflationary selection of the theory. 
Minimality is impressively consistent with the present observational
 bounds and it might be tested with the aid of low-energy soft
 parameters obtained in future experiments.
}
\end{center}
\end{titlepage}

\setcounter{page}{2}

\section{Introduction}

The minimal supergravity (mSUGRA)~\cite{Nilles:1983ge} is a very
interesting framework, since it has definite predictions on 
low-energy physics, which are well consistent with
the present observations. 
In particular, the absence of flavor-changing neutral currents (FCNC's)
in the present experimental precisions is one of the predictions of mSUGRA. 
If mSUGRA is (approximately%
\footnote{
We use the term mSUGRA in an approximate sense and
do not mean the strictly minimal K{\" a}hler potential
in this paper.})
realized in truth,
we wonder the reason of its selection by Nature,
since it seems no more symmetric
than nonminimal supergravity in the presence of a nontrivial
superpotential, as is the case for the standard model
of elementary particles.

In this paper, we point out that supergravity effective theory with a large
cutoff scale $M_*\,\, (>\!M_G)$ may be chosen by inflationary
selection of background vacuum structures~\cite{Iza},
which implies a specific type of mSUGRA theory. 
Here, $M_G$ is the reduced Planck scale, $M_G \simeq 2.4 \times
10^{18}$GeV. 
The large cutoff suggests relatively small gaugino masses,
which in turn indicate masses of squarks as large as a few TeV.
In this parameter region,
the mass of the lightest Higgs boson is easily raised
up to the current experimental limit. 
In spite of the large stop mass,
we may naturally obtain the breaking scale of electroweak symmetry
at ${\cal O}(100)$GeV~\cite{Barbieri:1987fn,Chan:1998xv}
due to renormalization group (RG) focus point behavior~\cite{Feng:1999zg}
of a supersymmetry (SUSY) breaking soft mass of a Higgs boson.

The large-cutoff theories might be realized in various corners of theory
moduli space, or (string
\cite{Bou,Smo}) landscape.
We do not specify concrete construction of such theories but simply
assume their presence.
The task in this paper is not to achieve constructive realization
of the large
cutoff but to seek a plausible way to select it
among vast possibilities on the landscape.%
\footnote{
See the conclusion in Ref.\cite{Izawa}.} 

The rest of the paper goes as follows.
In the next section, we consider possible inflationary selection of
minimality in supergravity. 
In section~\ref{sec:minimality}, we specify plausible boundary conditions
on gravity mediation of SUSY breaking. 
In section~\ref{sec:phen}, low-energy phenomenology is investigated by
means of RG analysis. 
Section~\ref{sec:conclusion} is devoted to conclusions and discussion.

\section{Possible Inflationary Selection}
\label{sec:inflation}

We are led by the following question:
what is expected beyond the standard model%
\footnote{
We suppose the standard model as a prerequisite
with its presently measured values of the couplings.}
as a typical structure of the natural laws?
We here consider inflationary selection
of background vacuum structures~\cite{Iza} and dwell on mediocrity
principle, which may prefer flatter inflaton potential~\cite{Smo,Vil}. 

For concreteness of presentation, let us adopt a simplest case of
supergravity inflation model~\cite{Yan,Izaw} as an example.
Namely, we consider a single-superfield model for slow-roll inflation.
In terms of a single chiral superfield $\f$, an inflaton $\v$ can be
provided by $\sqrt{2}$ times the real part of its lowest component.
We adopt a natural superpotential%
\footnote{This form is protected by nonrenormalization
or $R$ symmetry.}
\begin{equation}
 W = v^2 \f - {y \o n+1} \f^{n+1}
 \label{ADDEQ}
\end{equation}
and a generic K\"ahler potential
\begin{equation}
 K = |\f|^2 + {\k \o 4}|\f|^4 + \cdots,
\end{equation}
where $v^2,y,\k>0$ and the ellipsis denotes higher-order terms, which
may be disregarded. 
Here and henceforth in this section, we have taken the unit with a
cutoff scale $M_*$ equal to one.
Note that the small scale $v^2$ can be generated
dynamically~\cite{Yan,Hot}.  

The potential for the lowest component $\f$ is given in supergravity by 
\begin{equation}
  V = \exp\left({K \o M_G^2}\right)
      \left\{ \left( {\q^2 K \o \q \f \q \f^\dagger} \right)^{-1} |DW|^2
      - 3\left|{W \o M_G}\right|^2 \right\},
 \label{EPOT}
\end{equation}
where we have defined
\begin{equation}
 DW \equiv {\q W \o \q \f} + {\q K \o \q \f}{W \o M_G^2}.
\end{equation}
Thus, the potential of the real part $\v$ is approximately given by
\begin{equation}
 V(\v) = v^4 - {\k \o 2}v^4\v^2
              - {\l \o n!}\v^n
 \label{POT}
\end{equation}
for $n \geq 3$ and $0 < \l, v^2, \v \ll 1$ with
${\l / n!} \equiv {y v^2 / 2^{{n \o 2}-1}}$.
The parameters $n, v,$ and $\l$ are potentially under control by
symmetry.  
Let us fix them hereafter, for simplicity of argument.

We adopt slow-roll approximation~\cite{Lyt}.
The slow-roll inflationary regime is prescribed by the condition
\begin{equation}
 \e(\v) = {1 \o 2} \left(M_G{V'(\v) \o V(\v)} \right)^2 \leq 1,
 \quad |\y(\v)| \leq 1,
 \label{COND}
\end{equation}
where
\begin{equation}
 \y(\v) = M_G^2{V''(\v) \o V(\v)}.
\end{equation}
For the potential Eq.(\ref{POT}), we obtain
\begin{equation}
\begin{array}{l}
 \displaystyle
 {\e \o M_G^2} \simeq {1 \o 2}
          \left({-\k v^4\v - {\l \o {(n-1)!}}\v^{n-1} \o v^4}
            \right)^2
        = {\v^2 \o 2}\left(\k + {\l v^{-4} \o {(n-1)!}}\v^{n-2}\right)^2, \\
 \noalign{\vskip 2ex}
 \displaystyle
 {\y \o M_G^2} \simeq {-\k v^4 - {\l \o {(n-2)!}}\v^{n-2} \o v^4}
        = -\k - {\l v^{-4} \o {(n-2)!}}\v^{n-2},
\end{array}
\end{equation}
as slow-roll parameters.%
\footnote{
Thus the slow-roll condition Eq.(\ref{COND}) is satisfied for $\v \leq \v_f$
where 
\beq
 \v_f^{n-2} \simeq {(n-2)!(1-\k M_G^2) \o \l v^{-4}M_G^2},
 \nonumber
\eeq
which provides the value $\v_f$ of the inflaton field at the end of
inflation. 
An initial value $\v_i$ of the inflaton field amounts to the
corresponding number $N$ of total $e$-folding as
\cite{Izaw} 
\beq
 N = \int_{\v_f}^{\v_i} \! d\v \, M_G^{-2}{V(\v) \o V'(\v)}
 \simeq {1 \o (n-2)\k M_G^2} \ln \left\{ {1-\k M_G^2 \o 1+(n-2)\k M_G^2}
 \left(1+{(n-1)! \k \o \l v^{-4} \v_i^{n-2}}\right) \right\}.
 \nonumber
\eeq
}
These parameters, $\epsilon$ and $|\eta|$, characterize the flatness of
the inflaton potential.
From a viewpoint of mediocrity principle, some tuning for flatter
inflaton potential may be favored.  
For flatter potential, total amount of inflation becomes larger and
inflation lasts longer, even possibly turns out to be eternal, to result
in larger volume of habitable universe.  

Small parameters $\e$ and $|\y|$ are achieved by tuning two apparent
factors for flatter inflaton potential. 
One is obviously the coupling $\k$, which is small for small $\e$ and
$|\y|$. 
The other is the reduced Planck scale $M_G$, which is also small for
small $\e$ and $|\y|$, provided radiative corrections due to
gravitational interaction controlled by $M_G$ are loop suppressed and
affect the effective coupling $\k$ by at most order unity for $M_G 
\simeq M_*/4\pi$. 
This latter case is assumed in the following discussion,%
\footnote{
This seems possible in view of potential quantization~\cite{Bag} of
Newton's constant in supergravity, though the size of the radiative
corrections may be sensitive to the structure of ultraviolet
physics~\cite{Gai} and beyond the scope of effective theory approach.} 
which corresponds to mSUGRA with a large cutoff $M_*$ compared to the
gravitational scale $M_G$. 
It leads to a particular pattern of effective Lagrangian parameters,
whose details will be given in the following sections.

In the remainder of this section, let us further see possible
implications of the large cutoff $M_* \simeq 4\pi M_G$ on inflationary
selection of background vacua. 
For that purpose, we consider multiple succession of inflations with
each inflationary stage naturally preparing the initial conditions for
the next stage~\cite{Kaw}. 
The background vacua with multiple inflations seem to constitute
remarkable ingredients in inflationary selection,
based on which we seek a typical structure of the natural laws.

The point is that the tuning of the scale $M_G$ might simultaneously
realize successive inflations which are favorable according to
mediocrity principle. 
This is to be contrasted to multiple tunings of each coupling ($\k$ in
the above example) corresponding to each inflaton potential.

In the theory (moduli) space, the background vacua
({\it i.e.} particular theories)
with small K{\" a}hler couplings and/or small gravitational scale may induce
inflation and
be realized in Nature, which is at the heart of the inflationary selection.

\section{Plausible Minimality}
\label{sec:minimality}

Motivated by the discussion in the previous section, we assume a large
cutoff $M_*$ at an input scale $Q_0$, below which we adopt the
RG equations of the minimal supersymmetric standard model. 
This hypothesis leads to mSUGRA theory, since the large cutoff
suppresses higher dimensional operators in the K\"ahler potential.   
Before we explore low-energy implications of our large-cutoff hypothesis
in the next section, let us set more detailed RG boundary conditions at
the input scale%
\footnote{
We utilize the input scale $Q_0=M_G$ or $Q_0=M_{\rm GUT}$
on occasion postulating that
the ultraviolet contributions of RG above the so-called GUT scale
$M_{\rm GUT} \simeq 2\times 10^{16}$~GeV are not significant.}  
around $M_G$ for gravity mediation of SUSY breaking.

As usual, the minimal K\"ahler potential generates a
universal soft SUSY breaking mass for chiral multiplets $m_{\rm
scalar}=m_0$.    
The universal scalar mass $m_0$ results in the universality of the
scalar masses of squarks and sleptons in the first two generations
providing a solution to the FCNC problem. 

This minimality will be tested in the next generation accelerator
experiments by examining spectra of the squarks and sleptons for
$m_0<{\cal O}(10)$TeV. Hence we restrict our attention to this range
in this paper. 
In order to attain sizable gaugino masses $M_i~(i=1,2,3)$, we adopt a
singlet chiral superfield $Z$ (Polonyi field)~\cite{Din} with its $F$
term as the dominant SUSY breaking source in the hidden sector of
gravity mediation. 
Note that the SUSY breaking scale can be generated dynamically without
its cosmological problem~\cite{Hot,IY}. 
Then we obtain $M_i = {\cal O}(M_G/M_{*}) m_{3/2}$ through the $F$ term of
\begin{eqnarray}
 \frac{f_i}{M_*}Z{\cal W}_{i\alpha}{\cal W}_i^{\alpha},
\end{eqnarray}
where ${\cal W}_{i\alpha}$ and $f_i~(i=1,2,3)$ denote field strength
chiral superfields for gauge multiplets and their order one
coefficients, respectively, and $Z$ has a SUSY-breaking $F$ term as $F_Z
\simeq \sqrt{3}m_{3/2} M_G$. 
Furthermore, the universal scalar mass $m_0$ can be expressed in terms
of the gravitino mass $m_{3/2}$ as $m_0 = m_{3/2}$. 
Namely, for $M_* \simeq 4 \pi M_G$ advocated in the previous section,
the spectrum of the supersymmetric standard model (SSM) particles
has a hierarchical structure, $m_0\gg |M_{i}|$. 
For $m_0=m_{3/2} < {\cal O}(1)$TeV, gaugino masses of ${\cal
O}(M_G/M_*) m_{3/2}$ would be too small and thus we are led to adopt
an mSUGRA boundary condition $m_0 = {\cal O}(1-10)$TeV.

In addition to $m_0$ and $M_i$, the Polonyi field $Z$ in mSUGRA also
determines so-called $A$ parameters of SUSY breaking.
The $A$ parameter for each (scalar)$^3$ coupling is proportional to a
universal $A_0$ parameter and the corresponding Yukawa coupling
constant for the minimal K\"ahler potential.
Since $A_0$ is proportional to the vacuum expectation value (VEV) of the
Polonyi field $Z$, that is, $A_0 \simeq (\langle Z^* \rangle/M_G)
\,m_{3/2}$, the assumption $|A_0| \lsim m_0$ is not so implausible within a
well-controlled expansion on $Z/M_G$ in supergravity effective theory.%
\footnote{ 
The potential in supergravity has the $e^{K/M_G^2}$ factor, which forces
$|\vev{Z}| \lsim M_G$ provided $K\simeq |Z|^2$.
Note that such a range of the $A_0$ parameter is adequate to avoid
color symmetry breaking. 
}

So far, we have concentrated on the SUSY breaking parameters in mSUGRA.  
By virtue of the Polonyi field, we also naturally obtain 
the supersymmetric Higgs mixing parameter $\mu$
of the electroweak order:
the $\mu$ term can be provided through the Giudice-Masiero (GM)
term~\cite{Giudice:1988yz} in the K\"ahler potential 
\begin{eqnarray}
 K \supset \frac{{\cal O}(1)}{M_*}Z^* H_u H_d,
\label{eq:GM}
\end{eqnarray}  
which relates $\mu$ to the SUSY breaking parameters,
where $H_{u,d}$ denote the up-type and down-type Higgs superfields.
Then the parameter $\mu$ is
expected to be of the same order of the gaugino masses: $\mu\sim
M_i\simeq {\cal O}(M_G/M_*) m_{3/2}$. 
This GM term yields a specific form of the so-called $B$ term,%
\footnote{
Our convention for the Higgs mass parameters is given in the Appendix.
}
which is given explicitly in section~\ref{sec:Bterm}.

\section{Low-Energy Phenomenology}
\label{sec:phen}

In the previous section, we have proposed specific mSUGRA boundary
conditions: $m_0 = {\cal O}(1-10)$TeV\
with a hierarchical structure%
\footnote{
This hierarchy is natural in the large-cutoff theory
with $M_* \simeq 4\pi M_G$,
whereas it requires at least ${\cal O}(M_G^2/M_*^2)$ tuning
(implemented by some flavor symmetry) in the ordinary mSUGRA theory
with $M_G$ as a cutoff scale, even if we presuppose minimality
to put aside the corresponding tuning.}
$m_0\gg |M_i|, |\mu|$.
It is nontrivial that the present scenario admits the electroweak 
symmetry breaking at the correct energy scale, in particular,
since masses of squarks, sleptons, and Higgs bosons are very large
at the input scale $Q_0$.
Fortunately, the mSUGRA boundary conditions
turn out to be consistent with the present observational constraints on
the electroweak physics.
In this section, we show numerical analyses which indicate
this consistency. 
For the sake of explanatory convenience,
let us consider the case $\tan\b \gsim 10$ in the followings,
though numerical estimates include 
results on the case with smaller $\tan \b$
(see Fig.\ref{fig:MAPlow}).
Here, $\tan\b \equiv v_u/v_d$ is the ratio of the two VEV's
of the neutral Higgs fields, $v_{u,d} \equiv \vev{H^0_{u,d}}$.

\subsection{Consistency of the electroweak symmetry breaking}
\label{sec:EWSB}

With large $\tan\b$,
the scale of electroweak symmetry breaking
tends to be controlled by the value of a SUSY breaking soft mass
parameter $m_{H_u}^2$ (see Eq.(\ref{eq:EWSB1}) below)
at a relevant scale for the electroweak physics,
to be called the electroweak scale.

In mSUGRA, the running value of the parameter $m_{H_u}^2$
is related to the original parameters
$m_0, M_i$, and $A_0$ by
\begin{eqnarray}
 m_{H_u}^2 = a m_0^2 + b^i |M_i|^2 + c |A_0|^2 + d^i Re(M_i A_0^*), 
\label{eq:mHufit}
\end{eqnarray}
where the coefficients $a - d^i$ are scale-dependent functions of
dimensionless gauge and Yukawa
coupling constants.%
\footnote{
Approximate analytical expressions for the coefficients $a - d^i$ can be 
found in Ref.\cite{Cod}, for instance.
}
As discussed in Refs.\cite{Barbieri:1987fn,Chan:1998xv,Feng:1999zg},
the coefficients $a$ and $c$ are of
order $10^{-2} - 10^{-1}$ at the electroweak scale for
the pole mass of the top quark $m_{t}\simeq 170-180$~GeV,
which results in the RG focus point behavior.

The smallness of the coefficient $a$ comes from a cancellation between
$m_{H_u}^2(Q_0) = m_0^2$ at the input scale $Q_0$ and RG contributions
at the renormalization scale $Q$:
\begin{eqnarray}
& &\delta a(Q) m_0^2 \simeq
-\frac{1}{2}\bar{m}^2_0 \cdot (1-e^{L(Q)}) 
\simeq -\frac{1}{3} \bar{m}^2_0;
\nonumber \\
& &\qquad \bar{m}^2_0 \equiv
(m_{H_u}^2+m_{Q_3}^2+m_{U_3}^2)(Q_0) = 3 m_0^2,
\label{eq:Qdep} \\
& &\qquad L(Q) = \frac{1}{16 \pi^2}\int^{\ln Q}_{\ln Q_0}
12|y_t|^2(Q') d \ln Q',
\nonumber
\end{eqnarray}
where the subscripts $Q_3$ and $U_3$ represent
an $SU(2)_L$ doublet quark and a singlet 
up-type quark, respectively, in the third family, and $y_t$ denotes the
top Yukawa coupling constant.%
\footnote{ 
Here, we assume that the bottom Yukawa coupling constant $y_b$ is
negligible compared to $y_t$. 
Even if $y_b \simeq y_t$, the $m_0$ insensitivity of $m_{H_u}^2$ is
valid, although the RG contribution from the
bottom-type squark becomes important~\cite{Feng:1999zg}.
}
In the first equation above, we use the fact $e^{L(Q)}\simeq 1/3$ at
the electroweak scale for $m_t \simeq 170$~GeV.

On the other hand, the smallness of the coefficient $c$ can be traced to
the RG evolutions of (scalar)$^3$ coupling constants $A_i \cdot
y_i$~($i=t,b,\tau$). 
As the renormalization scale is lowered from the input scale $Q_0$,
the $A_i$'s
become small exponentially from $A_0$ and the contributions from the $A$
terms to the RG evolution of $m_{H_u}^2$ become small.
Thus, the $A_0$ dependence of $m_{H_u}^2$ at the electroweak scale
({\it i.e.} the coefficient $c$) is relatively small.

As a result, $m_{H_u}^2$ is suppressed compared to $m_0$ at the
electroweak scale for $|M_i| \ll m_0$, $|A_0| < m_0$
and becomes of the same order of the gaugino masses.

\begin{figure}
 \begin{center}
  \begin{minipage}{0.49\linewidth}
   \begin{center}
   (a) $m_{t} = 174$~GeV
    \includegraphics[width=.95\linewidth]{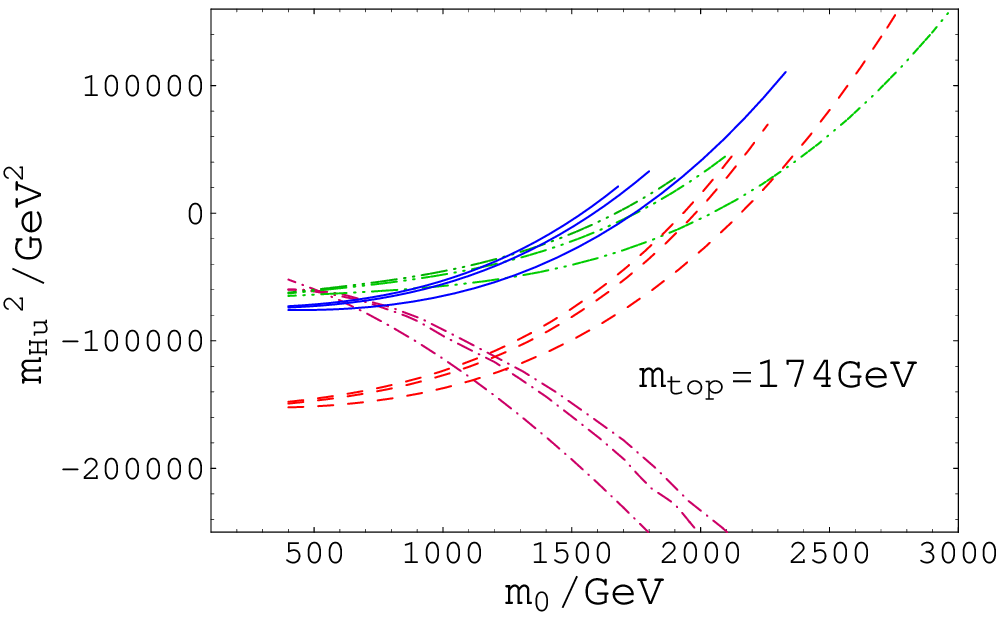}
   \end{center}
  \end{minipage}
  \begin{minipage}{0.49\linewidth}
   \begin{center}
   (b) $m_t = 178$~GeV
    \includegraphics[width=.95\linewidth]{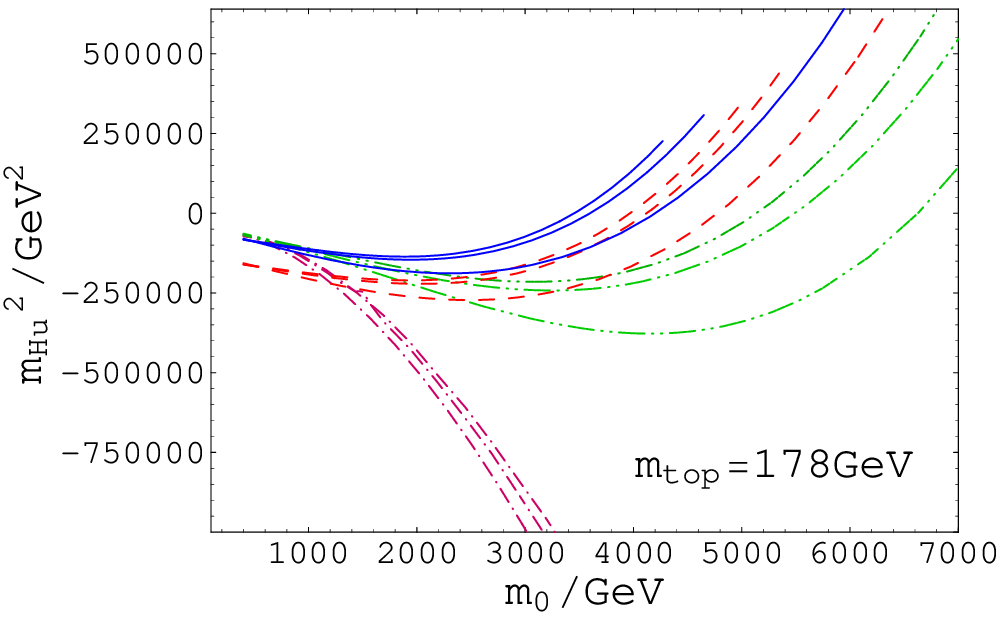}
   \end{center}
  \end{minipage}
 \end{center}
 \caption{
The $m_0$ dependence of $m_{H_u}^2$ 
at the stop mass scale $Q=Q_{\tilde t}$
for sgn$(\mu)=+1$ and for the top quark pole mass 
(a) $m_{t} = 174$GeV or (b) $m_t = 178$GeV. 
The blue (solid) lines correspond to $M_{1/2}= 200$GeV and $A_0 = 0$GeV, 
the red (dashed) lines $M_{1/2}= 300$GeV and $A_0 = 0$GeV,
the green (dash-dot-dotted) lines $M_{1/2}= 200$GeV and
$A_0 = 0.5m_0$,
and the purple (dash-dotted) lines $M_{1/2}= 200$GeV and $A_0 = m_0$.
The three lines of each type correspond to $\tan\beta = 10,20$, and
 $30$, respectively, from below.
} 
 \label{fig:mHu}
\end{figure}

In Fig.\ref{fig:mHu}, we show the $m_0$ dependence of $m_{H_u}^2$ at a
typical stop mass scale
$Q = Q_{\tilde{t}} =
(m_{\tilde{t}_1}m_{\tilde{t}_2})^{1/2}$.
In this computation, we impose
the boundary condition $m_{scalar}^2 = m_0^2$ at the
GUT scale ({\it i.e.} $Q_0=M_{\rm GUT}$),%
\footnote{ 
We assume that
the change of the input scale from $M_{\rm GUT}$ to $M_G$ does not
disturb the hierarchy $|m^2_{H_u}(Q_{\tilde{t}})|\ll m_0^2$. 
For instance, this is the case in the grand unification scenario,
since ${\bf 10} \supset (Q_3,U_3)$   
and ${\bf 5} \supset (H_u)$ have the same RG trajectory between the
$M_{\rm G}$ and $M_{\rm GUT}$ in the limit of $M_i$ vanishing.
}
and take a universal gaugino mass $M_i = M_{1/2}$, for simplicity.
As expected, the value of $m_{H_u}^2$ is much suppressed compared to
the corresponding $m_0^2$ for the case of $|M_i|\ll m_0$
and $|A_0|< m_0$.
We have plotted it for two central values of the observed 
top quark pole mass $m_t$: one is extracted from the Particle Data
Group~\cite{PDG2004} $m_t = 174.3\pm 5.1$~GeV and the other from the
recent CDF and D0 results~\cite{Azzi:2004rc} $m_t =178.0\pm 4.3$~GeV.  

In the SSM, the parameter $\mu$ is related to the $Z^0$ boson
mass $m_{Z^0}$ by
minimizing the effective Higgs potential, and it can be expressed at the
tree level as
\begin{eqnarray}
 \frac{1}{2} m_{Z^0}^2
 = \frac{m_{H_d}^2-m_{H_u}^2 \tan^2\beta}{\tan^2\beta-1}
 - |\mu|^2
 \simeq {m_{H_d}^2 \o \tan^2\beta}-m_{H_u}^2 - |\mu|^2.
\label{eq:EWSB1}
\end{eqnarray}
Thus, $m_{H_u}^2$ of the order of the gaugino masses
manages to generate
the electroweak symmetry breaking at the correct energy scale,
or $m_{Z^0} = 91.2$~GeV,
with the $\mu$ parameter naturally implied by the GM term (\ref{eq:GM}).
Here, the parameters in Eq.(\ref{eq:EWSB1}) are regarded to be values
at the electroweak scale,
while the RG evolution of $\mu$ from the electroweak scale to the input
scale is negligible in order estimation (see Eq.(\ref{eq:RGEmu})
in the Appendix).

\begin{figure}
 \begin{center}
  \begin{minipage}{0.49\linewidth}
   \begin{center}
    (a) $m_{t} = 174$~GeV
    \includegraphics[width=.95\linewidth]{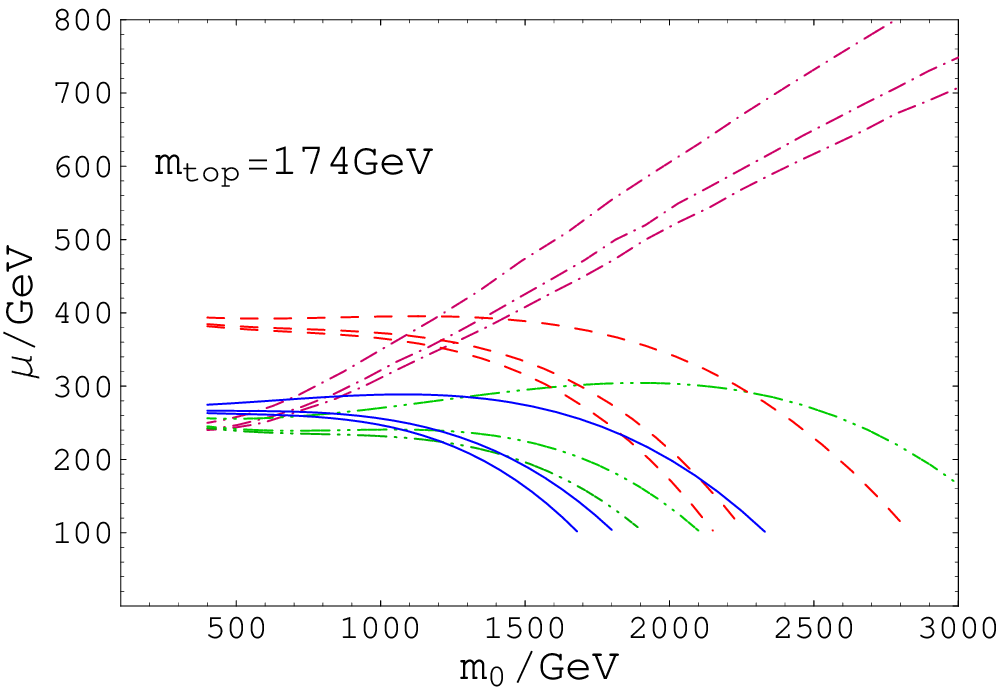}
   \end{center}
  \end{minipage}
  \begin{minipage}{0.49\linewidth}
   \begin{center}
    (b) $m_t = 178$~GeV
    \includegraphics[width=.95\linewidth]{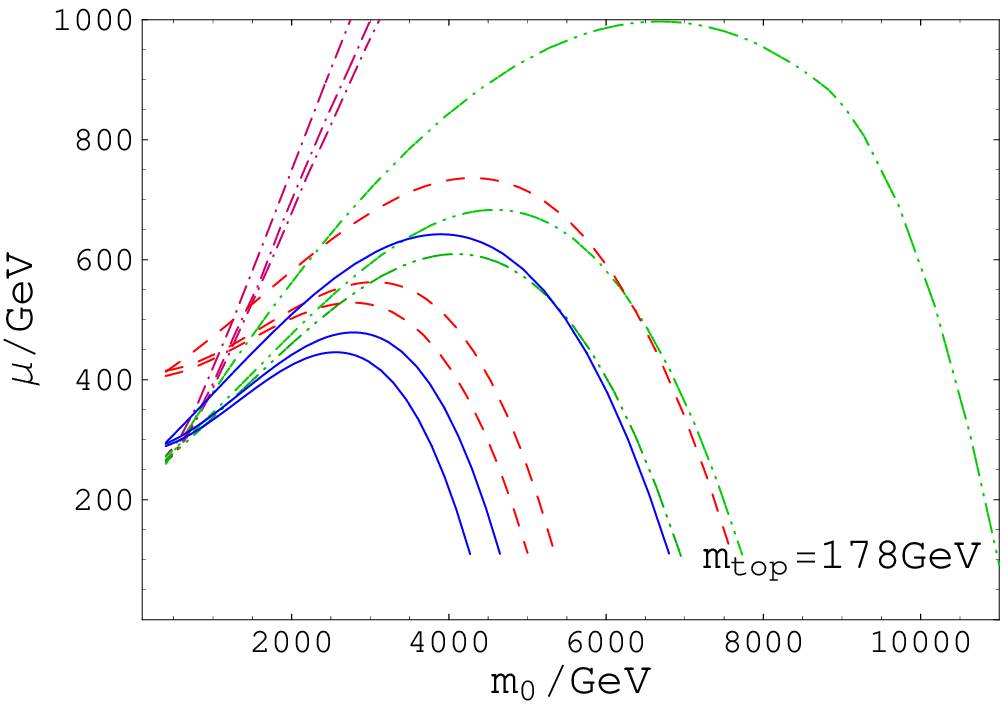}
   \end{center}
  \end{minipage}
 \end{center}
 \caption{
 The $m_0$ dependence of $\mu$ at the stop mass scale $Q=Q_{\tilde t}$
 for sgn$(\mu)=+1$ and for the top quark pole mass 
 (a) $m_{t} = 174$GeV or (b) $m_t = 178$GeV. 
 The notations for the lines are the same as in
 Fig.\ref{fig:mHu}, except for the
 corresponding values of $\tan\beta$ reversed in order:
 $\tan\beta = 10,20$, and $30$ from above.
}
 \label{fig:mu}
\end{figure}

In Fig.\ref{fig:mu}, we show admissible values of the $\mu$ parameter which
yield the observed value of $m_{Z^0} = 91.2$~GeV.  
To determine the value of $\mu$, we have used the
ISAJET~7.69 code~\cite{Paige:2003mg},
which takes into account the one-loop corrections
to the effective Higgs potential and the two-loop RG evolutions of
parameters.%
\footnote{
The discrepancy of the value of the $\mu$ parameter among computational 
codes is discussed in Ref.\cite{Allanach:2003jw}.
}
The minimization of the effective Higgs potential is also performed
at the typical stop mass scale $Q = Q_{\tilde{t}}$.
We see that the value of $\mu$ should be much suppressed compared to 
$m_0$ for $|M_i|\ll m_0$, $|A_0|< m_0$ and hence it is consistent
with our large-cutoff hypothesis.

Let us comment on the falling-off behavior
of the allowed $\mu$ in the very large $m_0$ region in
Fig.\ref{fig:mu}.    
This behavior stems from a large cancellation between the 
suppressed $m_0^2$ contribution and
the remaining ones
to the value of the $\mu$ parameter.
In such a region, the $\mu$ becomes very small due to the cancellation.

From a cosmological point of view, the parameter regions with tiny $\mu$
may provide a natural explanation for the observed dark matter
density~\cite{Feng:2000gh},   
since the lightest neutralino
is a bino-Higgsino mixture for such regions and its
relic abundance is in a cosmologically interesting range.

\subsection{Consistency of the tree-level $B$ term}
\label{sec:Bterm}

We have assumed the GM term as the origin of
the $\mu$ parameter of the electroweak order.
Then, the SUSY-breaking Higgs mixing parameter $B_0$ is related to
the $A_0$ parameter at the input scale.
For the term (\ref{eq:GM}),
the tree-level relation is given by~\cite{Giudice:1988yz}
\begin{eqnarray}
 B_0=B_0^{GM} \equiv \frac{2A_0 - 3 m_{3/2}}{A_0-3 m_{3/2}} m_{3/2}.
\label{eq:B}
\end{eqnarray}
In this subsection, we examine how this condition is satisfied
in the electroweak physics.%
\footnote{
A similar
analysis is performed in Ref.\cite{Ellis:2004qe}
for small $m_0$ regions.  
}

To study the matching condition Eq.(\ref{eq:B}), we take the following
procedure:
We first fix sampling values of
$(m_0,M_i,A_0,{\rm sgn}(\mu),\tan\b)$, and determine the required
values of $\mu$ and $B$ that reproduce $m_{Z^0} = 91.2$~GeV.
In addition to Eq.(\ref{eq:EWSB1}), we have a relation
\begin{eqnarray}
 B \mu = \frac{\sin 2\beta}{2}(m_{H_u}^2+m_{H_d}^2+2 |\mu|^2),
\label{eq:EWSB2}
\end{eqnarray}
which is also obtained by minimizing the tree-level
effective Higgs potential.
By means of Eqs.(\ref{eq:EWSB1}) and (\ref{eq:EWSB2}),%
\footnote{More precisely, their one-loop corrections are taken into
account in the numerical analysis below.}
we can obtain $\mu$ 
and $B$ at the electroweak scale for the given mSUGRA parameters.
Then, from the value of $B$ at the electroweak scale,
we compute the $B$ parameter at the input scale
({\it i.e.} $B_0$) and compare it with
$B_0^{GM}$ in Eq.(\ref{eq:B}) to see whether or not the
condition~Eq.(\ref{eq:B}) is satisfied.

We find from Eq.(\ref{eq:EWSB2}) that the required
value of $B$ at the electroweak scale is given by
\begin{eqnarray}
 B \sim \frac{m_0^2}{\mu\tan\beta},
\label{eq:Bapp}
\end{eqnarray}
since $m_{H_d}^2\sim m_0^2 \gg |m_{H_u}^2|,\, |\mu|^2$
and $\sin 2\beta/ 2\simeq
1/\tan\beta$ for $\tan\beta\gsim 10$. 

\begin{figure}
 \begin{center}
  \begin{minipage}{0.49\linewidth}
   \begin{center}
    (a) $m_{t} = 174$~GeV
    \includegraphics[width=.95\linewidth]{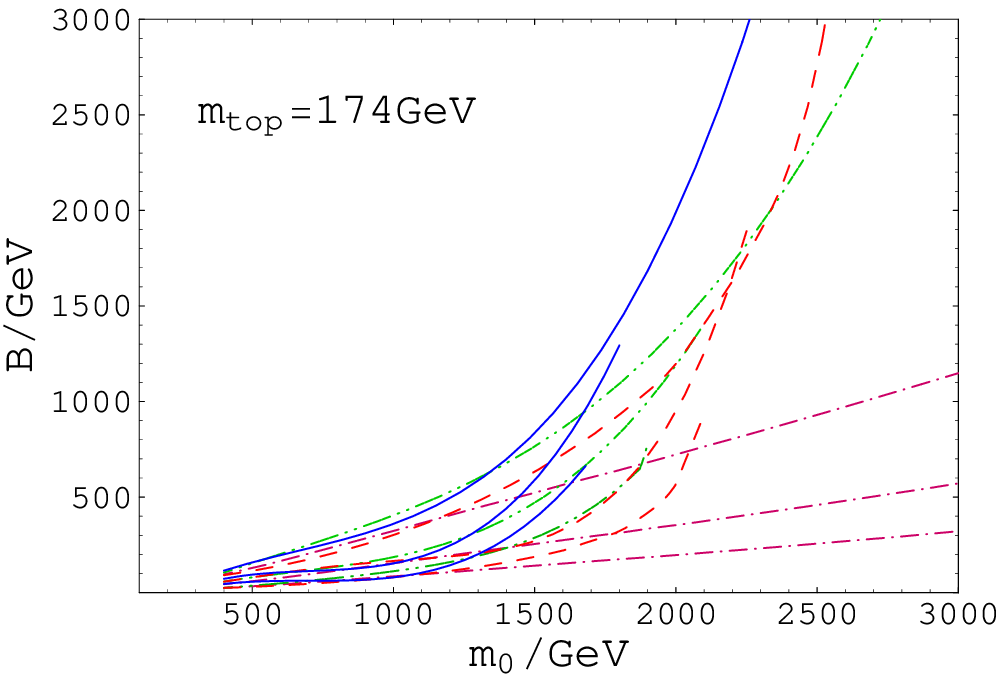}
   \end{center}
  \end{minipage}
  \begin{minipage}{0.49\linewidth}
   \begin{center}
    (b) $m_t = 178$~GeV
    \includegraphics[width=.95\linewidth]{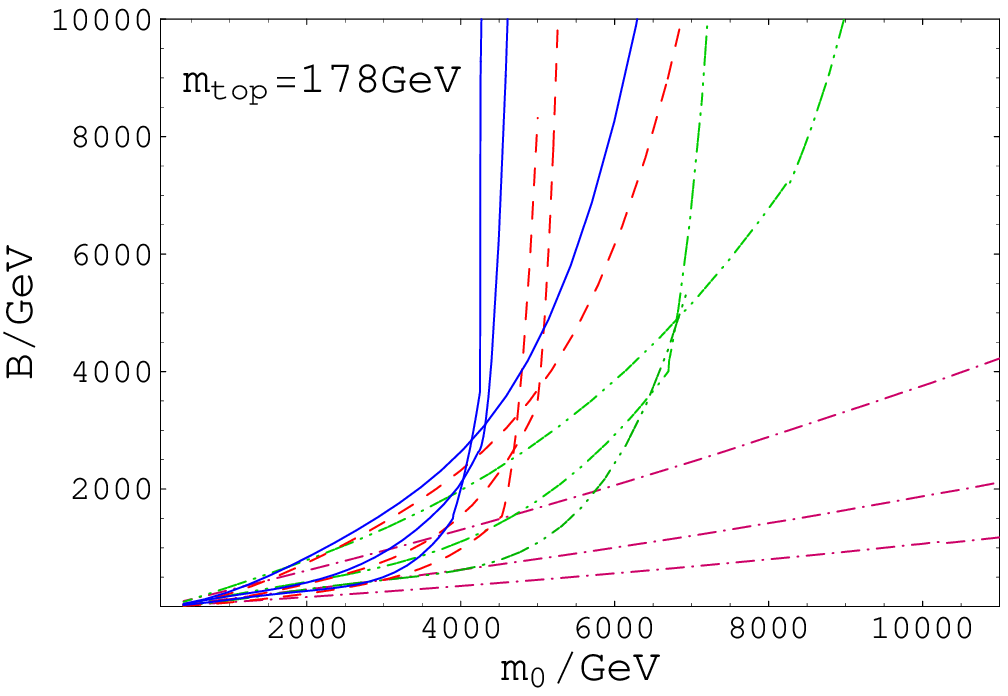}
   \end{center}
  \end{minipage}
 \end{center}
 \caption{ The $m_0$ dependence of $B$ at the stop mass
 scale $Q=Q_{\tilde t}$
 for sgn$(\mu)=+1$ and for the top quark pole mass
 (a) $m_{t} = 174$GeV or (b) $m_t = 178$GeV. 
 The notations for the lines are the same as in Fig.\ref{fig:mHu}.}
 \label{fig:B}
\end{figure}

In Fig.\ref{fig:B},
we show numerical results
on the value of $B$ at the
stop mass scale $Q=Q_{\tilde t}$ as a function of $m_0$. 
By comparing it with the $\mu$ parameter in Fig.\ref{fig:mu}, we find
that the value of $B$ becomes very large when the value of $\mu$ becomes  
very small, as expected
from Eq.(\ref{eq:Bapp}).

On the other hand,
from the relation~Eq.(\ref{eq:B}),
we see $B_0^{GM}\sim m_0$ for $|A_0|< m_0 = m_{3/2}$. 
Thus Eq.(\ref{eq:Bapp})
implies that the condition~Eq.(\ref{eq:B}) can be satisfied
for $\mu \sim m_0/\tan\beta \ll m_0$, 
since the RG evolution of $B$ from the electroweak scale to
the input scale is not significant: the RG equation of $B$ is
controlled by relatively small $M_i$ and $A$ parameter
contributions (see Eq.(\ref{eq:RGEs}) in the Appendix),
and thus $B\sim B_0$.  
As a result, we find that the condition~Eq.(\ref{eq:B}) is satisfied in a
certain parameter region.  

\begin{figure}
 \begin{center}
  \begin{minipage}{0.45\linewidth}
   \begin{center}
    (a) tan$\b = 20$ \\
    \includegraphics[width=.95\linewidth]{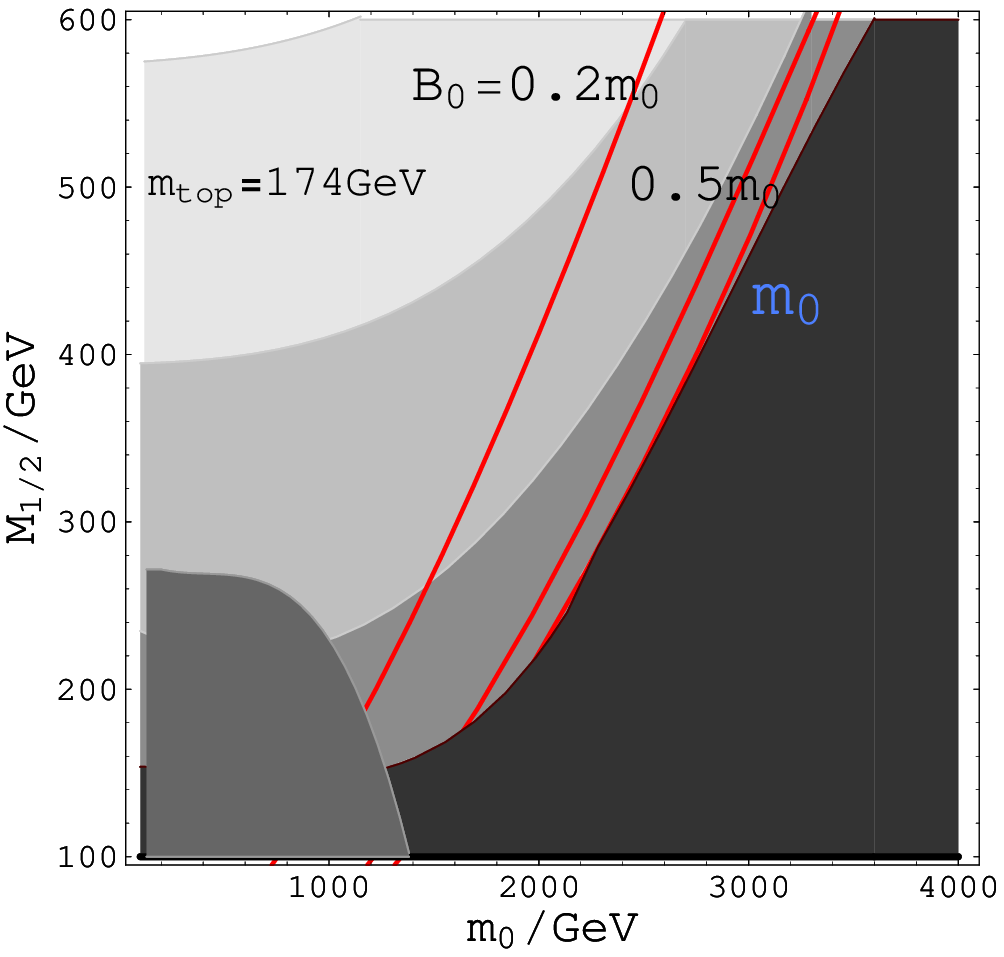}
   \end{center}
  \end{minipage}
  \begin{minipage}{0.45\linewidth}
   \begin{center}
    (b) tan$\b = 20$ \\
    \includegraphics[width=.95\linewidth]{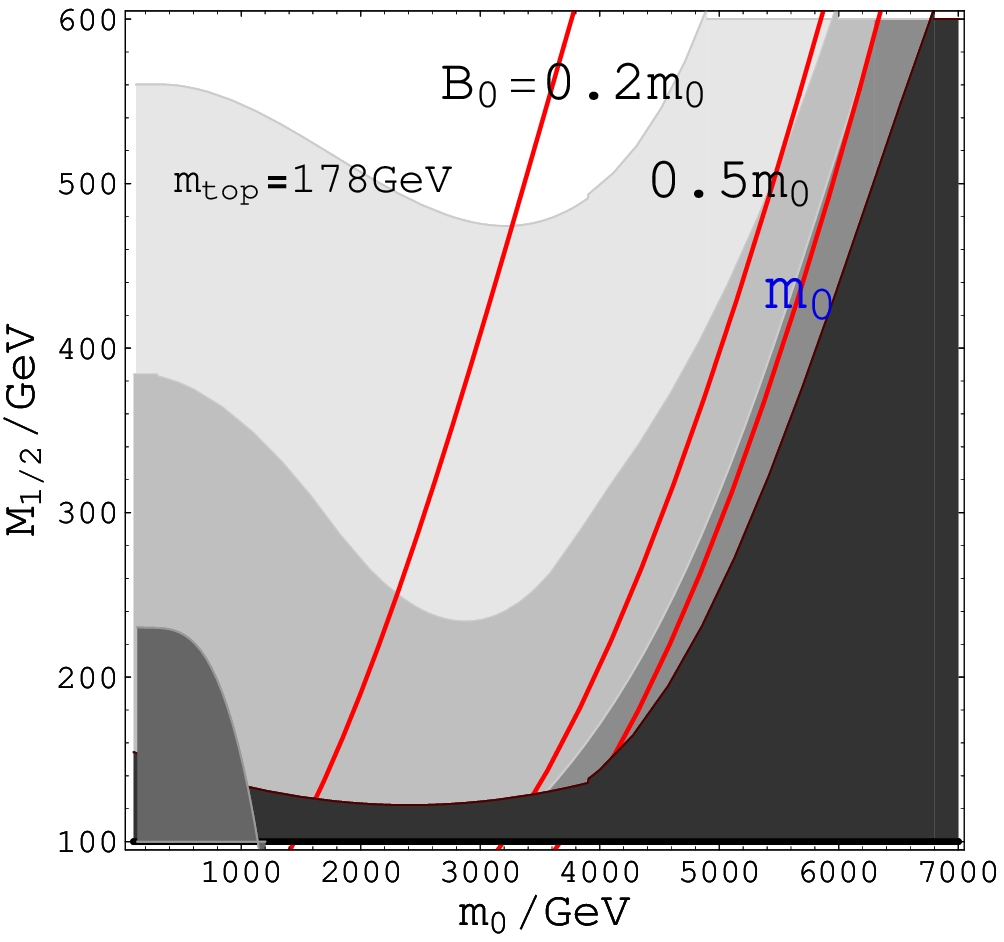}
   \end{center}
  \end{minipage}
 \end{center}
 \begin{center}
  \begin{minipage}{0.45\linewidth}
   \begin{center}
    (c) tan$\b = 10$ \\
    \includegraphics[width=.95\linewidth]{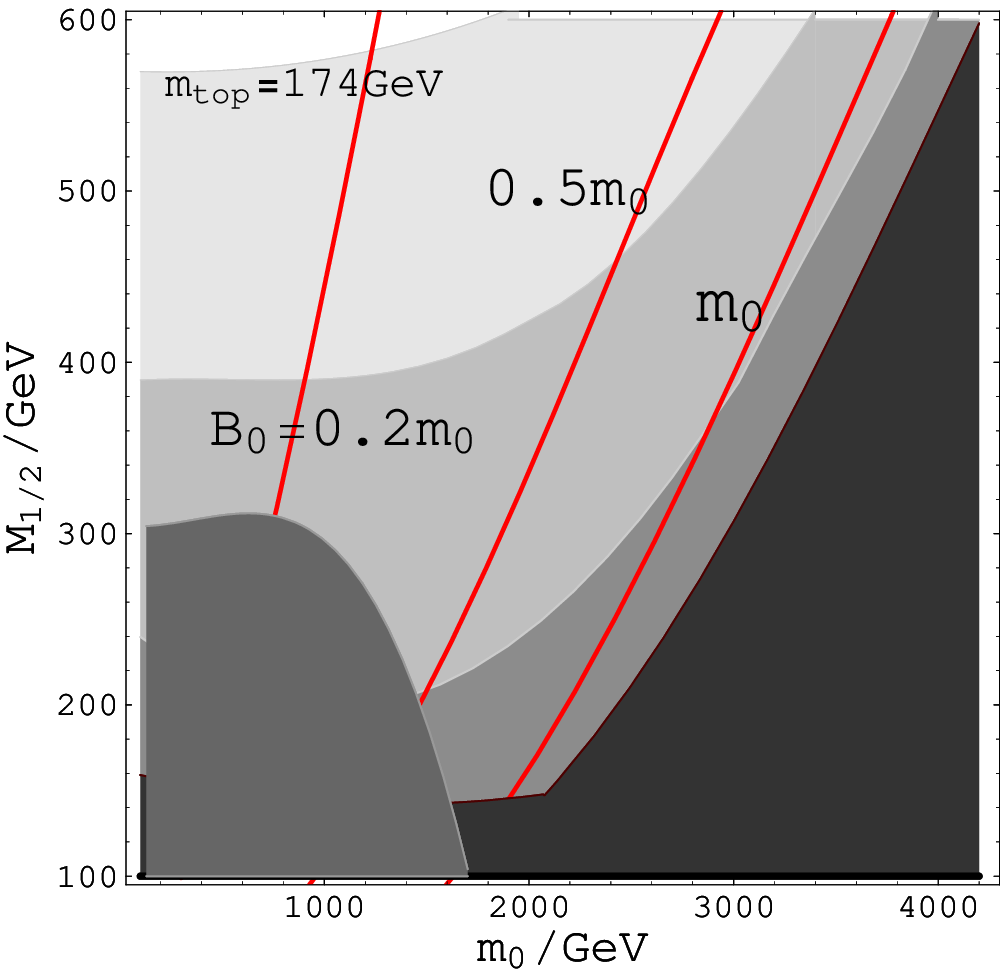}
   \end{center}
  \end{minipage}
  \begin{minipage}{0.45\linewidth}
   \begin{center}
    (d) tan$\b = 10$ \\
    \includegraphics[width=.95\linewidth]{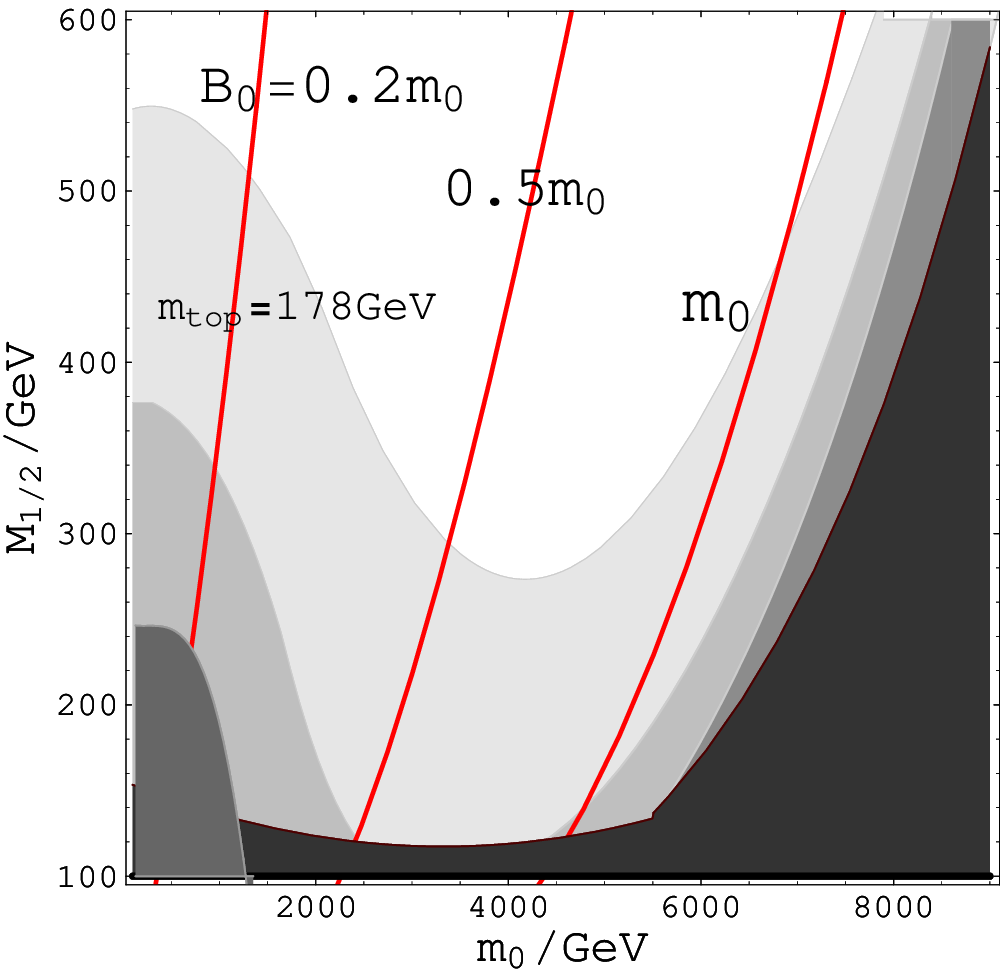}
   \end{center}
  \end{minipage}
 \end{center}
 \caption{Contour plots for the required values of $B=B_0$ at the input
 scale $Q_0 = M_{GUT} \simeq 2 \times 10^{16}$GeV
 in the ($m_0$, $M_{1/2}$) plane
 for the top quark pole mass (a,c) $m_{t} = 174$GeV or (b,d) 
$m_t = 178$GeV
and for (a,b) $\tan \beta = 20$ or (c,d) $\tan \beta = 10$.
In all the panels, we have fixed $A_0=0$, sgn$(\m)=+1$, and
the red (solid) lines correspond to $B_0 = 0.2 m_0, 0.5 m_0, m_0$,
respectively, from the left to the right. 
The GM term implies the parameter regions on the $B_0 = B_0^{GM}=m_0$ lines
(see Eq.(\ref{eq:EWSB2})).
The gray shaded regions correspond to the parameters where the required
 value of $\mu$ is smaller than $700$GeV, $500$GeV, and $300$GeV,
 respectively, from above.
The dark shaded
 regions at the bottom-left represent the Higgs mass bound
 $114$~GeV\cite{Higgs}, and
the black shaded regions are excluded by the chargino mass
 limit, $m_{\chi^{\pm}}\ge 104$GeV~\cite{chargino}.
 }
 \label{fig:MAP}
\end{figure}

\begin{figure}
 \begin{center}
  \begin{minipage}{0.45\linewidth}
   \begin{center}
    (a) tan$\b = 5$ \\
    \includegraphics[width=.95\linewidth]{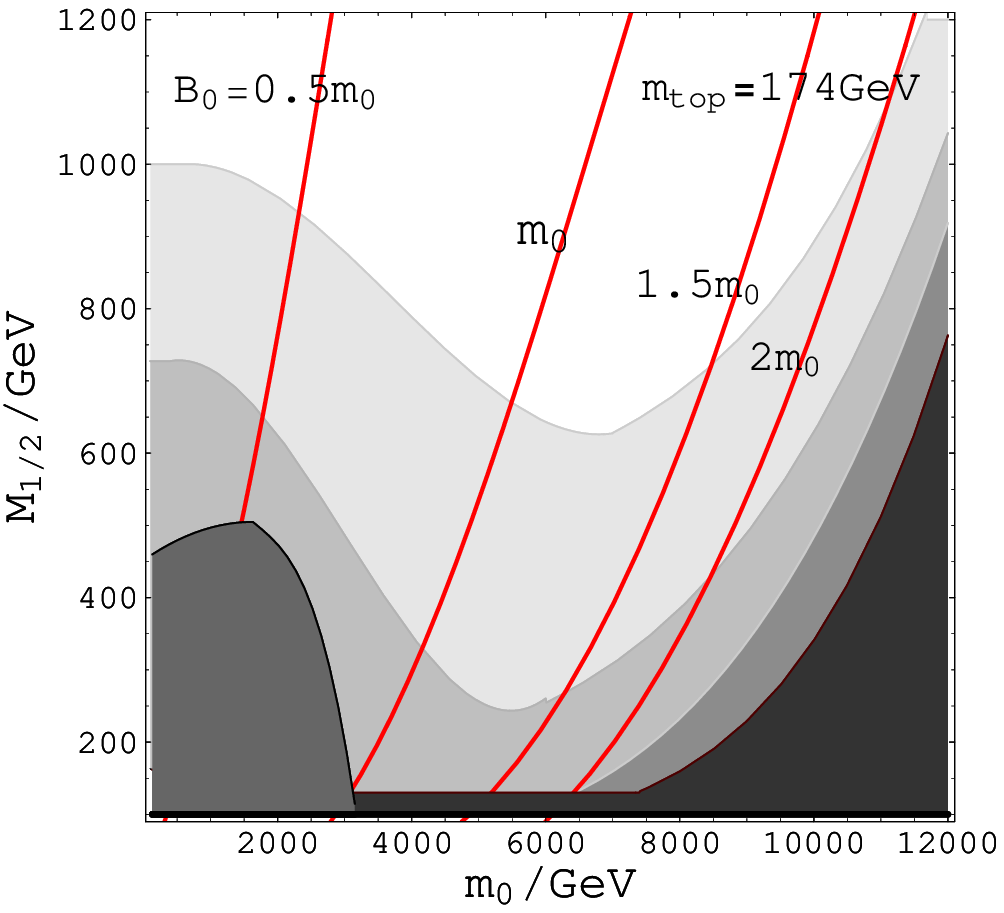}
   \end{center}
  \end{minipage}
  \begin{minipage}{0.45\linewidth}
   \begin{center}
    (b) tan$\b = 5$ \\
    \includegraphics[width=.95\linewidth]{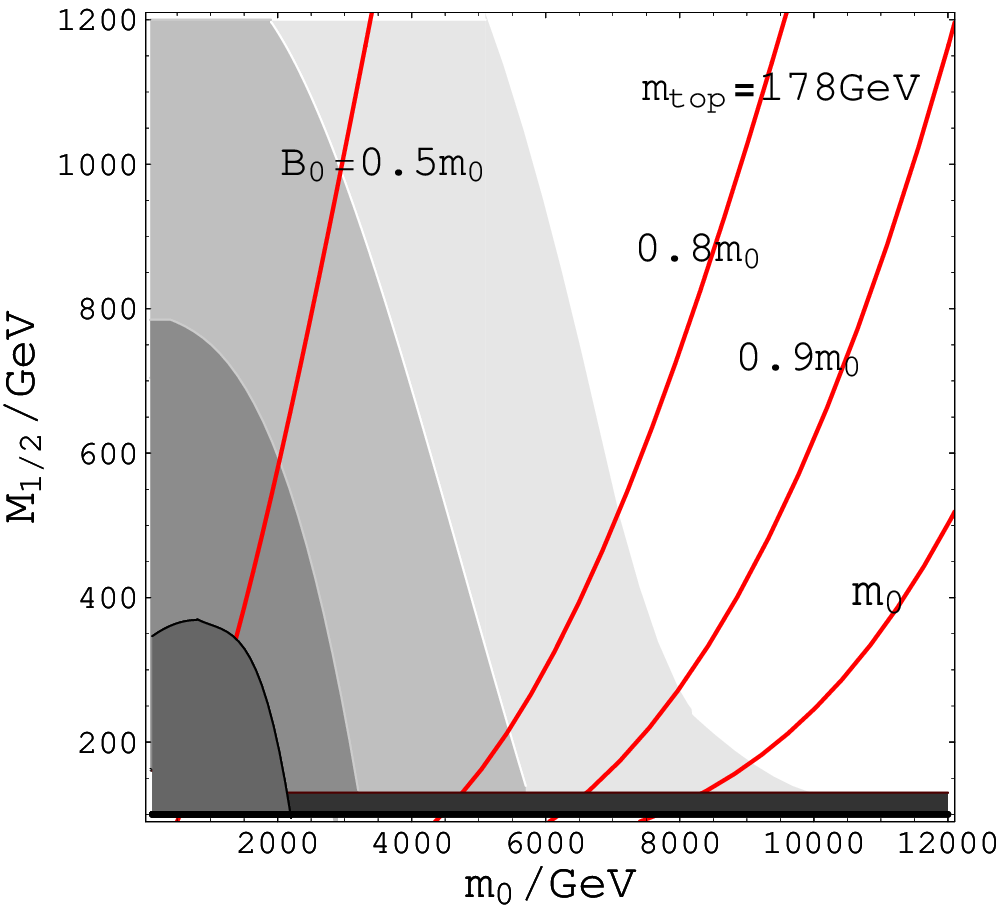}
   \end{center}
  \end{minipage}
 \end{center}
 \begin{center}
  \begin{minipage}{0.45\linewidth}
   \begin{center}
    (c) tan$\b = 4$ \\
    \includegraphics[width=.95\linewidth]{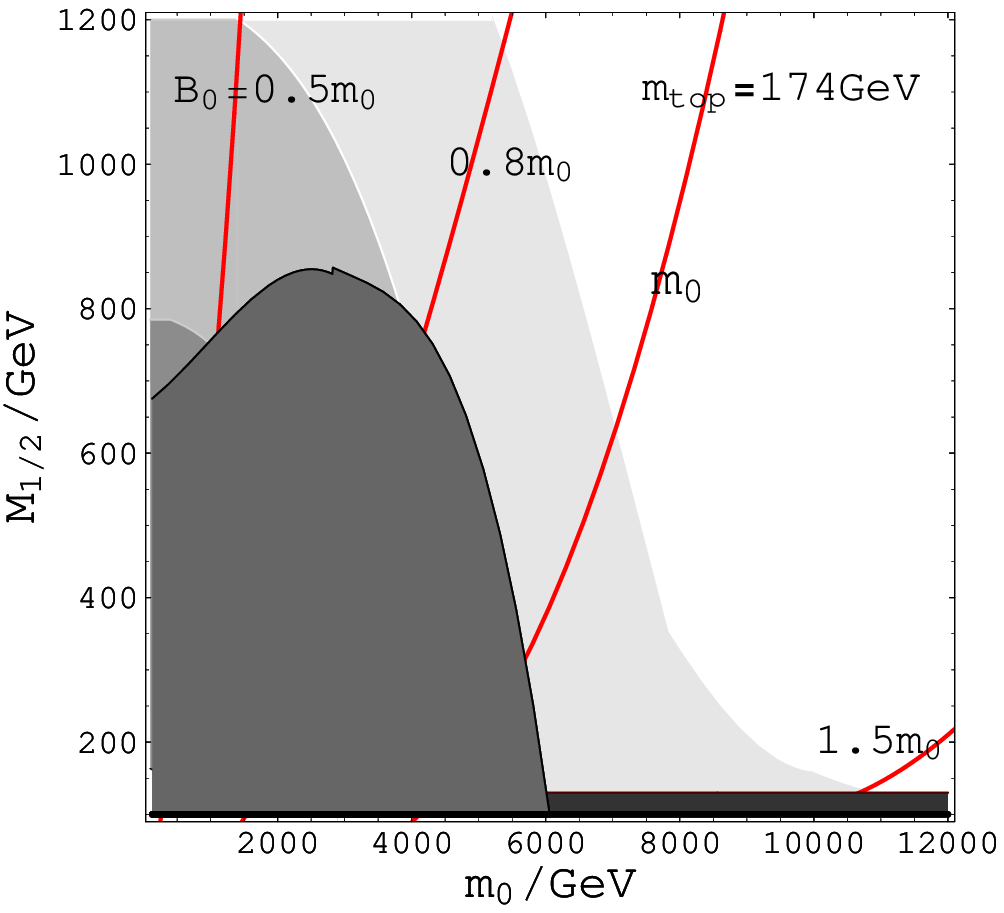}
   \end{center}
  \end{minipage}
  \begin{minipage}{0.45\linewidth}
   \begin{center}
    (d) tan$\b = 4$ \\
    \includegraphics[width=.95\linewidth]{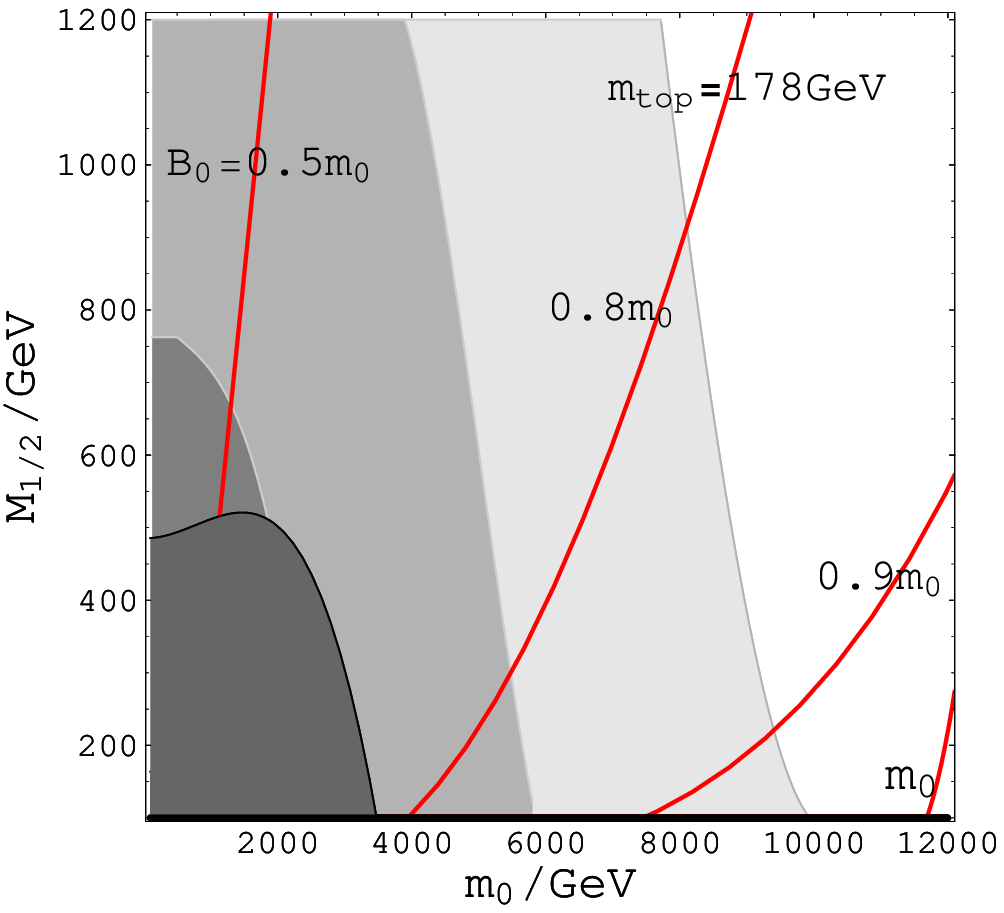}
   \end{center}
  \end{minipage}
 \end{center}
 \caption{Contour plots for the required values of $B=B_0$ at the input
 scale $Q_0 = M_{GUT} \simeq 2 \times 10^{16}$GeV
 in the ($m_0$, $M_{1/2}$) plane
 for the top quark pole mass (a,c) $m_{t} = 174$GeV or (b,d) 
$m_t = 178$GeV
and for (a,b) $\tan \beta = 5$ or (c,d) $\tan \beta = 4$.
In all the panels, we have fixed $A_0=0$, sgn$(\m)=+1$, and
the red (solid) lines correspond to $B_0$ fixed as indicated. 
The GM term implies the parameter regions on the $B_0 = B_0^{GM}=m_0$ lines
(see Eq.(\ref{eq:EWSB2})).
The gray shaded regions correspond to the parameters where the required
 value of $\mu$ is smaller than $1200$GeV, $900$GeV, and $600$GeV
 from above in the panel (a); smaller than $2000$GeV,
 $1500$GeV, and $1000$GeV from the right in the panels (b,c); and
 smaller than  $3000$GeV, $2000$GeV, and $1000$GeV from the right
 in the panel (d). 
 The dark shaded
 regions at the bottom-left represent the Higgs mass bound
 $114$GeV~\cite{Higgs}, and
 the black shaded regions are excluded by the chargino mass
 limit, $m_{\chi^{\pm}}\ge 104$GeV~\cite{chargino}.
 }
 \label{fig:MAPlow}
\end{figure}

In Figs.\ref{fig:MAP} and \ref{fig:MAPlow}, we plot
the contours of the values of $B_0$ at
the input scale $Q_0=M_{GUT}$
on the $(m_0,M_{1/2}\geq 0)$ plane with the fixed values of
$A_0 = 0$, sign$(\mu)=+1$, and $\tan\beta = 20,10,5,4$ as demonstrations.    
In the figures, we also show the value of $\mu$ at the stop mass scale
$Q=Q_{\tilde t}$.

Before examining the contours of $B_0$,
let us first understand the behavior of $\mu ={\rm constant}$~lines
in the figures.
Generic behavior of the $\mu = {\rm constant}$~lines
may be seen from the panels (b,d) in Fig.\ref{fig:MAP}
and (a) in Fig.\ref{fig:MAPlow}.
The lines are elliptic in small $m_0$ regions
(elliptic domains)
and hyperbolic in large $m_0$ regions
(hyperbolic domains).
They are parabolic in between
(parabolic domains),
where the value of $\mu$ is relatively insensitive to the variation of
$m_0$ with $M_{1/2}$ fixed.
In this perspective, the panels (a,c) in Fig.\ref{fig:MAP}
belong to hyperbolic domains
(continued from parabolic domains), where the $\mu = {\rm constant}$~lines
are hyperbolae
(in the large $\tan \b$ case
in Ref.\cite{Chan:1998xv});
and the panels (b,c,d) in Fig.\ref{fig:MAPlow}
belong to elliptic domains
(continued on parabolic domains), where the $\mu = {\rm constant}$~lines
are ellipses
(in the small $\tan \b$ case
in Ref.\cite{Chan:1998xv}).
We note that such behavior
may be obtained from Eqs.(\ref{eq:mHufit}), (\ref{eq:Qdep}),
and (\ref{eq:EWSB1})
with the stop mass as the renormalization scale%
\footnote{
In contrast, the RG focus point behavior
manifests itself
under the $m_0$ independent choice of the renormalization scale
$Q$ with the focus point given by $a(Q)=0$ in Eq.(\ref{eq:mHufit})
\cite{Feng:1999zg}.}
$Q=Q_{\tilde t} \sim m_0$.

Now, by inspection of $B_0=B_0^{GM}=m_0$~lines
due to the GM term (\ref{eq:GM}) for $A_0=0$ (see Eq.(\ref{eq:EWSB2}))
in the figures,
it is apparent that the large-cutoff theories with $m_0 \gg |M_{1/2}|, |\mu|$
correspond to those lines
along the hyperbolic domains in Fig.\ref{fig:MAP}
with moderate $\tan \b$.
In particular, the lower ends of the $B_0=m_0$ lines
are in small $M_{1/2}$ regions with $m_0$ of a few TeV,
which may be within the reach of
accelerator experiments in the near future.

More generally, the figures imply that, in the large-cutoff theory,
the tree-level relation%
\footnote{
This tree-level relation may suffer from possible corrections
of order $M_G/M_*$ at the input scale, 
which, we hope, is to be compared with future experimental results.}
of the GM term (\ref{eq:GM}) in mSUGRA
can be satisfied for various parameters
with $m_0\gg |M_i|,|\mu|$,
which are consistent with the present observational bounds.
Hence we conclude that the GM term works very well
with our large-cutoff hypothesis.

\section{Conclusions and Discussion}
\label{sec:conclusion}

Gravity-mediated supersymmetry breaking and primordial inflation are
expected to open windows into the Planck-scale physics through
observations on superpartner spectra~\cite{Nilles:1983ge} and on
temperature fluctuations of cosmic microwave background
radiation~\cite{Lyt}.  
In this paper, we have discussed realization of the large-cutoff theory
in supergravity, which is possibly selected through inflationary
dynamics. 
This large-cutoff hypothesis implies an mSUGRA
spectrum with a large hierarchy
between the universal scalar mass and the gaugino masses,
$m_0 \gg |M_i|$.  
Very encouragingly, despite of relatively large masses of scalar
particles, the electroweak symmetry breaking
can occur at the correct energy scale
with $m_0\gg |\mu|$
in phenomenologically viable parameter regions.   

In the large cutoff hypothesis, the absence of the FCNC process is
automatic, since all of the corresponding
higher dimensional operators in the K\"ahler  
potential are suppressed by the large cutoff $M_*$. 
In addition, with the current chargino mass bound, the hierarchical
spectrum $m_0\gg |M_i|,|\mu|$ predicts heavy sfermions at a few TeV, and
hence the CP problem in the SSM is ameliorated.     
In most of the parameter region
($m_0\gg |M_i|,|\mu|$) we are interested in,
the lightest supersymmetric particle 
is a neutralino, which is a good candidate for the dark matter
(see also the remark at the end of section~\ref{sec:EWSB}).

Finally, let us comment on the origin of matters in the universe in the
present scenario. The hierarchical spectrum implies
that the mass of the gravitino is of the order of a few TeV. 
In this case, the primordial abundance of the gravitino should be
suppressed not to disturb the Big-Bang Nucleosynthesis, which implies the
reheating temperature $T_R \lsim$~$10^{6-7}$~GeV~\cite{Kawasaki:2004qu}.  
Hence, the baryon asymmetry must be provided at the corresponding
low temperatures,
$T \lsim$~$10^{6-7}$~GeV.

\section*{Acknowledgements}

The authors wish to thank Y.~Sinbara for valuable discussion.
This work is partially supported by Grand-in-Aid Scientific Research (s)
14102004.

\section*{Appendix: Notation for the Higgs Potential}
\label{sec:notation}

In this appendix, we list our convention for the Higgs mass parameters.
We adopt the following form of the effective Higgs potential at
the tree level:
\begin{eqnarray}
 V &=& 
(|\mu|^2 + m_{H_u}^2) (|H_u^0|^2+|H_u^+|^2)
+(|\mu|^2 + m_{H_d}^2) (|H_d^0|^2+|H_d^+|^2) \nonumber\\
& & +B\mu \left\{(H_u^+H_d^- - H_u^0 H_d^0) + c.c.\right\}
+ \frac{1}{2} g^2 |H_u^+ H_d^{0*}+H_u^{0}H_d^{-*}|^2 \nonumber\\
& &+ \frac{1}{8}(g^2+g'^{2})(|H_u^0|^2+|H_u^+|^2-|H_d^0|^2-|H_d^+|^2)^2,
\label{eq:Higgspotential}
\end{eqnarray}
where the superscript $(0,+,-)$ of each field denotes its electric
charge, and $g$ and $g'$ denote the $SU(2)_W$ and $U(1)_Y$
gauge coupling constants, respectively.
With this convention, the RG equations for the $\mu$ and $B$ parameters are
given at the one-loop level by
\begin{eqnarray}
 \frac{d}{d \ln Q} \mu &=& \frac{\mu}{16\pi^2}
(3 |y_{t}|^2 + 3 |y_b|^2 + |y_\t|^2
 - 3 g^2 - g'^2),
\label{eq:RGEmu} \\
 \frac{d}{d \ln Q} B &=& \frac{1}{16\pi^2}
(6 A_t |y_{t}|^2 + 6 A_b |y_b|^2 + 2 A_\t|y_\t|^2 
+ 6 g^2 M_2 + 2g'^2 M_1),
\label{eq:RGEs}
\end{eqnarray}
where $A_i\cdot y_i$ $(i=t,b,\t)$ denotes the (scalar)$^3$ coupling constant
that is the supersymmetric counterpart of the Yukawa coupling $y_i$ for
each flavor $t, b$, or $\t$.


\end{document}